\documentclass[12pt]{article}
\usepackage{amssymb,amsmath,amscd,latexsym}
\newcommand{\be}{\begin{equation}}
\newcommand{\ee}{\end{equation}}
\newcommand{\bea}{\begin{eqnarray}}
\newcommand{\eea}{\end{eqnarray}}

\newcommand{\ba}{\begin{array}}
\newcommand{\ea}{\end{array}}

\newcommand{\bt}{\begin{tabular}}
\newcommand{\et}{\end{tabular}}

\newcommand{\fr}{\frac}
\newcommand{\ci}{\cite}
\newcommand{\cl}{\centerline}
\newcommand{\bs}{\bigskip}
\newcommand{\vs}{\vspace}
\newcommand{\hs}{\hspace}

\newcommand{\en}{\eqno}

\newcommand{\fnt}{\footnotetext}
\newcommand{\fm}{\footnotemark}
\newcommand{\bbib}{\begin{thebibliography}}
\newcommand{\ebib}{\end{thebibliography}}

\newcommand{\bm}{\boldmath}
\newcommand{\mbb}{\mathbb}
\newcommand{\mf}{\mathfrak}
\newcommand{\lra}{\longrightarrow}

\begin{document}
\titlepage
\hs{5cm}{\it LANDAU INSTITUTE preprint 29/05/97}

\vs{2cm}
\centerline{\bf ON TOPOLOGICAL  INTERPRETATION}

\centerline{\bf OF QUANTUM NUMBERS}

\bigskip

\centerline{\bf  S.A.Bulgadaev\fm{}}\fnt{This work is supported
by RFBR grants 96-02-17331 and 96-1596861}

\bigskip

\centerline{ Landau ITP RAS, 117334, Moscow,  Kosyghin street, 2}

\vs{1cm}

\cl{Abstract}

\vspace {1.0cm}

It is shown, how one can define vector topological charges for
topological exitations of
non-linear $\sigma$-models on compact homogeneous
spaces $T_G$ or $G/T_G$ (where $G$ is a simple
compact Lie group and $T_G$ is its maximal commutative
subgroup).
Explicit solutions for some cases, their energies and interaction of
different topological charges are  found.
A possibility of the topological interpretation of the quantum
numbers of groups and particles is discussed.

\newpage

\cl{\large {\bf Introduction}}

\bigskip

The discovery of new topologically stable solutions [1-4] has revived
the old, ascending to antiquity, hypothesis about a
topological nature of simplest particles. For example, Descartes offered
vortex model of magnetism \ci{5}, while lord Kelvin has putted forward a
conjecture
that atoms can be represented as knotted configurations
of "ether" \ci{6}. Though it has appeared that both conjectures are
non correct
they were very fruitful and have stimulated the
investigation of these phenomena as well as development of the vorticity
and of the knot theories \ci{7} and their applications in different regions
of physics from hydrodynamics till polymers [8-11].
An advantage of this idea in comparison with usual, considering simplest
particles as a pointlike, structureless objects, is that it offeres an
obvious image of elementary particles.
Later, analogous ideas have renewed periodically in connection with different
events in history of physics, (the most succes they achieved in condensed
matter physics, see for example \ci{11}), but, in the field theory,
they usually have been
based on one or
another topological invariant
and for this reason a set of topological charges, offered by them,
were not reach enough
for complete description of
real particles with internal symmetries (there is a huge number of literature
on this subject, we note here only \ci{12} for further information).

A modern physics has many common
properties with physics of second part of XIX century with condensate
of various fields playing
role of ether or others hypotetical liquids. For this reason a rebirth
of hypothesis about topological
origin of particles and their properties is very natural and it
looks attractive from  physical point of view, since it brings more deep
understanding of the nature of particles and gives more obvious interpretation
of their structure and quantum numbers in contrast with existing formal,
purely Lie-algebraic and structureless, description. To this one can add,
that  for this period a topology was strongly
developed and adequate mathematical methods for description and investigation
of various topological problems has been constructed (see for example \ci{13}).

In order to this interpretation was possible,  the
following points are necessary:

1) the corresponding field theory must
have degenerate vacuum,
which forms some manifold ${\cal M}$
with nontrivial topology,

2) a theory must have
solutions with nontrivial topology and finite (or logariphmic) energy,

3) the set of possible topological charges must be the same as
the set of possible quantum numbers.

The two first points are now fulfilled for such topological solutions as
solitons, vortices, instantons and monopoles, but a third point remains
incompleted, at least for first three exitations.

Usually quantum numbers of particles are determined by the weights
of irreducible representations to which they belong.
As is known the quantum numbers or weights of simple compact groups $G$
are connected with the
maximal commutative subgroup $T_G$ of $G$  - the maximal abelian Cartan
torus \ci{14}.

All possible weights
of compact simple group form $n$-dimensional lattice ${\mbb L}_w$,
where $n$ is
a rank of group $G.$ Thus, the topological charges must also belong to
the same lattice.
Earlier it had been shown that solitons with topological isovectorial
charges,
belonging to the corresponding root, weight and dual root lattices
${\mbb L}_r,{\mbb L}_w,{\mbb L}_v$ of  groups $G$ exist in
two-dimensional theories, generalizing sine-Gordon theory and connected
with characters of groups with rank $n > 1$ \ci{15}. These charges can be
related
with homotopy group $\pi_0$ of vacuum configurations of these models,
which form infinite discrete sets, coinciding with those lattices.

In general a possibility of existence of topological
exitations in
systems with degenerate vacuum,  depends on a non-triviality
of the homotopy groups $\pi_i({\cal M})$ of vacuum space ${\cal M}.$
For example, to usual vortices with integer
topological
charges $e\in {\mbb Z}$ and one-dimensional instantons \ci{16}  corresponds
$\pi_1(S^1) = {\mbb Z}$, while the two-dimensional instantons correspond to
$\pi_2(S^2) = {\mbb Z}$ \ci{4,13}.
As is shown in \ci{17}, for obtaining topological charges belonging
to weight lattice of compact
simple Lie group $G$ it is necessary
to consider in case of vortices a torus $T_G$ and in a case
of instantons a more special homogeneous space of $G$, a flag space
$F_G=G/T_G.$

In this paper we continue consideration of these manifolds ${\cal M}.$
It
will be shown how one can define corresponding vector topological charges and
some explicit solutions with such  topological charges will be found.
We will deal mainly with non-linear $\sigma$-models on manifolds
${\cal M},$ since they are effective theories for large class
of models having ${\cal M}$ as their vacuum manifolds.
We also discuss
%when compact homogeneous spaces
%${\cal M}=T_G$ and ${\cal M} = G/T_G = F_G$ can emerge as the vacuum
%manifolds
%of some field theories and
conditions under which quantum numbers of groups
admit topological interpretation.

\newpage

\cl{\large {\bf 1. Manifolds with vectorial  homotopy group $\pi_1$}}

\bs

The study of the manifolds with vectorial (i.e. with rank $n>1$)
homotopy group
it is convenient to begin with manifolds with vectorial $\pi_1({\cal M})$
groups. For our purposes it will be enough to consider only spaces with
abelian $\pi_1,$ since the corresponding topological charges must commute
with each other. We also confine ourselves by free homotopic groups for
simplicity.
Though their properties are simple and well known we need discus them with
a special attention paid to their vector structure, since it help us under
consideration of spaces with vectorial $\pi_2.$
The simplest generalization of circle
$$S^1 = e^{2\pi i\phi}, \quad 0\le \phi \le 1$$
is a  torus $T^n,$
which equals to the direct product  of $n$ circles
$$
T^n    = \bigotimes^n_{i=1} S^1_i.
\en(1)
$$
The torus $T^n$ can be also represented as a coset space of $n$-dimensional
euclidean space $R^n$
$$
T^n = {\mbb R}^n/{\mbb Z}^n,
\en(2)
$$
where ${\mbb Z}^n$ is $n$-dimensional simple cubic integer-valued
lattice
$$
{\mbb Z}^n = \bigoplus_{i=1}^n {\mbb Z}_i
\en(3)
$$
A more general $n$-dimensional torus $T_L$ of rank $n$ can be defined as
a factor
$$
T_L = {\mbb R}^n/{\mbb L}
\en(4)
$$
where ${\mbb L}$ is some full $n$-dimensional lattice in ${\mbb R}^n$
$$
{\mbb L} = \sum_{i=1}^n n_i {\bf e}_i, \,\, n_i \in {\mbb Z}_i, \,\,
{\bf e}_i \in \{{\bf e}_i\}_L,
\en(5)
$$
Here a set of linearly independent vectors  $\{{\bf e}_i\}_L, \,i=1,...,n$
forms a basis of lattice ${\mbb L}$.
For $T^n$ $\{{\bf e}_i\}_L$ coincides with the
canonical orthonormal basis
$${\bf e}_i = (0,..,0,1_i,0,...,0), \,i=1,...,n.$$
A basis of any lattice can be choosed by many ways, all of them are related
by modular transformations
$$
\{{\bf e}'_i\}_L = M \{{\bf e}_i\}_L, \, det(M) =\pm 1,
\en(6)
$$
where matrices $M$ have integer-valued entries and for basises having the
same orientation $det(M)=1.$ Usually the most convenient basis is the one
having vectors with minimal possible norms.

The  first homotopy group of torus $\pi_1(T^n)$, being the homotopical
mapping
classes of $S^1$ into $T^n,$ is usually written as
$$
\pi_1(T^n) = \bigoplus^n_{i = 1}{\mbb Z}_i = {\mbb Z}^n,
\en(7)
$$
where $i$-th component describes mappings of $S^1$ into $i$-th circle
of $T^n.$ It is clear that mapping from different components cannot
annihilate each other.
The same expression is often used for other tori $T_L.$
Then this form   indicates only
that coordinates of different homotopical classes are integer-valued
in some basis, but does not contain any information about
this basis.
It means that one must introduce in $\pi_1(T_L)$
a vectorial structure.

To find out an explicit form of this basis one can take into account
a quotient nature of $T_L.$ Then
it is desirable to choose it
compatible with the  vectorial structure of the covering space
${\mbb R}^n.$ The most natural way to do this is to conserve the euclidean
vectorial structure of ${\mbb R}^n$ and of ${\mbb L},$ embedded into it.
Then one comes to the next expression for $\pi_1(T_L)$
$$
\pi_1(T_L) = {\mbb L}
\en(8)
$$
where now ${\mbb L}$ contains in explicit form its basic
vectors ${\bf e}_i.$ This form can have more information
than (7), since, except of integer-valuedness of homotopical
classes ($n_i \in {\mbb Z}_i$), it contains geometrical characteristics
of $T_L.$ One need also to know  metric and corresponding scalar
product in space of topological charges, which may differ from that
of covering space (see below).
They will be very important for discussion of interaction of
topological exitations with different topological charges.

Since to each basic vector ${\bf e}_i$ corresponds an elementary nontrivial
homological cycle $\gamma_i$ of $T_L,$ the corresponding homological group
of cycles $H_1(T_L,{\mbb Z})$ of torus $T_L$ coincides with $\pi_1(T_L).$
This is a realization of the Gurevich theorem about isomorphism of first
non-trivial homotopic and homological groups \ci{13}.
Consequently, $H_1(T_l,{\mbb Z})$ also has vectorial structure,
which can be written as
$$
\mbox{\bm $\gamma$} = \sum n_i {\mbox{\bm $\gamma$}}_i, \,
{\mbox{\bm $\gamma$}}_i = \gamma_i {\bf e}_i, \,\, n_i \in Z.
\en(9)
$$
There is also a cohomological group of 1-forms $\delta \in H^1(T_L,{\mbb
R}),$
dual to $H_1(T_L,{\mbb Z}),$
with basis $\{{\mbox{\bm$\delta$}}_i\} = \{{\bf s}_id\phi^i\}$ and
a convolution
$$
({\mbox{\bm$\gamma$}}_i \circ {\mbox{\bm $\delta$}}_k) =
({\bf s}_i {\bf e}_k) \oint_{\gamma_i} d\phi^k =  \delta_{ik}
\en(10)
$$
where $\{{\bf s}_i\}$ is a basis of dual lattice ${\mbb L}^*$ and
$$
({\bf s}_i {\bf e}_k) = \delta _{ik},
\en(11)
$$
here $()$ means an euclidean scalar product.

Usually the topological charges $Q$ of $T^n$ are defined as a set of
integer numbers, corresponding to the winding numbers around
elementary cycles $\gamma_i$,
$$
Q = (q_1,...,q_n), \ q_i \in {\mbb Z}_i
\en(12)
$$
The winding numbers can be written as integrals of 1-forms $\delta_k$,
dual to the cycles $\gamma_i$, in the parameter space of $T^n$
$$
q_i = \oint_{\gamma_i}d\phi =
\oint_{\gamma_i}\sum_{k=1}^n n_k d\phi_k = n_i,\,\,
d\phi = \sum_{k=1}^n n_k d\phi_k
$$
or
$$
q_i = \oint_{\gamma} d\phi_i =
\sum_{k=1}^n n_k \oint_{\gamma_k}d\phi_i = n_i, \,\,
\oint_{\gamma_i} d\phi_k = \delta_{ik},\,\,
\gamma = \sum_{k=1}^n n_k \gamma_k
\en(13)
$$
Here $\gamma$ and $d\phi$ are some closed contour and 1-form on $T^n$.
Analogous expressions for $Q$ are usually used for any $T_L$. In this
formal form all $Q$ have the similar, integer-valued, structure and
all geometrical properties are hidden again in abstract basises of
elementary cycles or 1-forms.

For obtaining vector topological charges, characterizing vectorial homotopic
group, one can use any vector structure,
mentioned above. For example, one can lift on the covering
space ${\mbb R}^n$
and define topological charges ${\bf Q}$
as line integrals of vector 1-forms $dx^k$  $({\bf x}\in {\mbb R}^n)$ along
preimage (or pull-back) curve
${\gamma}^{*}$ of the closed contour $\gamma$
$$
{\bf Q} =  \int_{\gamma^{*}} d{\bf x} =
\sum_{i=1}^n n_i {\bf e}_i
\en(14)
$$
The latter equality follows from the fact that both ends of curve
${\gamma}^{*}$ must belong to the lattice ${\mbb L}.$
Note that the same expression can be derived if one
use the vectorial structure from (9) in the space of
cycles $\gamma$
$$
{\bf Q}=\oint_{\mbox{\bm $\gamma$}} d\phi =
\sum n_i {\bf e}_i \oint_{\gamma_i} d\phi_k = \sum_{i=1}^n n_i {\bf e}_i
\en(15)
$$
Further we adopt a second approach, using vectorial structure of homology
group, to all considered ${\cal M},$ since they will satisfy to conditions of
the Gurevich theorem. This choice is very obvious and
usually the possible differential forms are fixed
by theory action or by symmetry properties.

Thus we have introduced vectorial topological charges of torus $T_L$,
containing in the
explicit form the lattice's basis $\{{\bf e}_i\}_L.$
For obtaining metric and corresponding scalar product in space of topological
charges one must consider concrete realization of torus $T_L$ and a chiral
model on it.
Torus $T_L$ of rank $n$ can be realized as diagonal $n$-dimensional
matrices ${\bf t}$
$$
{\bf t}= diag(e^{2\pi i(s_1 \phi)},..., e^{2\pi i(s_n \phi)})
\en(16)
$$
where $()$ denotes usual scalar product in $n$-dimensional euclidean
space ${\mbb R}^n,$ and a set $\{{\bf s}_i\}$ is a basis of dual lattice
${\mbb L}^{*}$ from (10-11).
It is easy to see that to each elementary cycles ${\bf e}_i$ corresponds
one exponent
in ${\bf t}$ with dual vector ${\bf s}_i.$

The lattice ${\mbb L}^{*}$ is a weight lattice of torus $T_L,$ considered
as a group.
For topological interpretation of all weights it is necessary that
${\mbb L}\supseteq {\mbb L}^{*}.$ For example,
a lattice $\mathbb{Z}^n$ coincides with its dual one, consequently,
all weights
of  group $T^n$
have  topological interpretation.

For this reason in a case of general tori $T_L$ an important role play tori,
corresponding
to the self-dual lattices ${\mbb L} = {\mbb L}^{*}$, since for them all weights
have one to one correspondence with topological charges. But it is not
enough for exact isomorphism between them, because one must have
the same metric in space of topological charges. As we will see below,
it doesn't  take place for tori $T_L.$

Now we pass to discussion of the compact chiral models,
connected with $T_L$
and will describe the explicit
realizations of vector topological charges introduced above.
The first model generalizes two-dimensional
nonlinear $\sigma$-model on a circle $S^1$ \ci{19}. We will use for $T_L$
a realization in the
form (16). One can define two-dimensional nonlinear $\sigma$-model on $T_L$
by next action ${\cal S}$ (or energy $E$)
$$
{\cal S} = \fr{1}{2\alpha} \int d^2x Tr({\bf t}^{-1}_{\mu} {\bf t}_{\mu})
=\fr{1}{2\alpha} \int d^2 x ( \phi_{\mu}\cdot \phi_{\mu}),
\en(17)
$$
where ${\bf t}_{\mu}= \partial_{\mu}{\bf t},\,
\phi_{\mu} =\partial_{\mu}\phi , \,\mu=1,2$
and here appears new scalar product $(\cdot)$, defined by effective
$n$-bein metric $g_{ik}$ on torus space,
$$
(\phi\cdot\phi)= \sum g_{ik} \phi^i \phi^k,\quad
g_{ik} = \sum_{a=1}^n s^a_i s^a_k
\en(18)
$$
It easy to see that metric $g_{ik}$ for general lattice ${\mbb L}$ does
not coincide
(except case of $T^n$) with euclidean one on ${\mbb R}^n.$ Exactly this metric
will define interaction
of different vector topological charges on torus $T_L.$

This chiral model can be considered as effective theory,
corresponding in long-wave limit to the Ginzburg-Landau type theories or
to discrete lattice type models with vacuum manifold ${\cal M}= T_L.$
Due to nontriviality of $\pi_1 (T_L),$ it must have vortex-type solutions.
These vortex solutions are ill-defined for small scales, where the full
action must be used for finding a structure of the vortex core.
But for description of the topological properties only long-wave
behavior will be important  and for this reason we shall use
at small distances instead of core a short-wave cut-off parameter $a,$
which is analog of the lattice constant or of the core radius.

The corresponding equations
$$
\partial^2 \phi_i =0
\en(19)
$$
due to its linearity
have $N$-"vortex" solutions in ${\mbb R}^2$ plane with $N$ punctures
$$
\mbox{\bm $\phi$}({\bf x}) = \sum_{i=1}^N {\bf q}_i\fr{1}{\pi}
\arctan(\fr{y-y_i}{x-x_i}),\,\,
{\bf q}_i \in {\mbb L},\,\,(q_i s_a)\in {\mbb Z}, \; (x,y)\in {\mbb R}^2
\en(20)
$$
Energy of one "vortex" with topological charge ${\bf q}=
\sum_{i=1}^n n_i {\bf e}_i$
is logariphmically divergent
$$
E= \fr{(q\cdot q)}{2\alpha} 2\pi \ln \fr{R}{a},\;
(q\cdot q) = \sum g_{ik} q^i q^k = \sum_{i=1}^n n_i^2,
\en(21)
$$
where $R$ is a space radius and $a$ is some small size cut-off.
It follows from (22) that independently on lattice ${\mbb L}$
topological charges,
corresponding to different cycles, do not interact as in the case of
torus $T^n!$  This is a consequence of appearence of effective metric $g_{ik}$
on tori space. Energy of $N$-"vortex" solution $E_N$ with whole topological
charge ${\bf Q} = \sum _{i=1}^N {\bf q}_i = 0$ is
$$
E_N= \fr{2\pi}{2\alpha}\sum_{i\ne k}^N ({\bf q}_i\cdot{\bf q}_k)
\ln \fr{|x_i-x_k|}{a} + C(a) \sum_i^N ({\bf q}_i\cdot {\bf q}_i),
\en(22)
$$
where scalar product of topological charges is
$$
({\bf q}_i\cdot{\bf q}_k)= \sum_{a=1}^n n^a_i n^a_k
\en(23)
$$
and $C(a)$ is some nonuniversal constant, determining "self-energy"
(or core energy)
of vortices and depending on type of core regularization.
Thus, we see that properties of vortex exitations of non-linear $\sigma$-model on all tori $T_L$
(with tori dimension equal to their rank, they also include complex abelian
tori) are similar to those of
non-linear  $\sigma$-model,
defined on tori $T^n.$  In particular, the  topological charges, corresponding
to different cycles, do not interact with each other though they belong to
lattice ${\mbb L}\ne Z^n$. It can be understood,
since on these tori one can always deform initial cycles into canonical ones.
At the same time this result confirms that we have introduced vector structure
correctly.

But this is not a whole story. There are tori different from $T^n$ and
$T_L$ types. They can be called degenerated ones because their rank $n$ is
smaller than their dimension $p.$  We consider them in the next section.

The second model we want to consider here is  one-dimensional
conformal $\sigma$-model on $T_L,$ which generalizes analogous $\sigma$-model
on $S^1.$ Its action is defined as
$$
{\cal S}= \fr{1}{2\alpha} \int dx dx'\fr{|{\bf t}(x)-{\bf t}(x')|^2}{(x-x)^2}=
\en(24)
$$
$$
\fr{1}{2\alpha}\int dx dx'\; 2\sum_1^n (1-\cos(s_i(\phi(x)-\phi(x')))
\en(25)
$$
where $x\in {\mbb R}^1.$
Following method of paper \ci{16}, one can show, that this model has
$N$-instanton solutions of the form (20)
$$
\mbox{\bm $\phi$}(x)=\sum_{i=1}^N {\bf q}_i \fr{1}{\pi}\arctan
\fr{x-a_{1i}}{a_{2i}},
\en(26)
$$
where $a_{1i}, a_{2i}$ are abitrary constants, characterizing place and width
of corresponding instanton, and all topological charges ${\bf q}_i$ in (26)
must satisfy simultaneously a condition
$$
(q_i s_k)\le 0 \quad \mbox{or} \quad (q_i s_k)\ge 0
\en(27)
$$
for all ${\bf s}_k, \, k=1,...,n.$
This condition is a generalization of analogous
condition in $\sigma$-model on $S^1,$ which, in its turn, is an analog
of analiticity (or anti-analiticity) property of two-dimensional instantons.
The corresponding action is
$$
{\cal S}_N = \fr{(2\pi)^2}{2\alpha} \sum_{i=1}^n |(s_i Q)|,\;
{\bf Q}= \sum_{i=1}^N {\bf q}_i
\en(28)
$$
Such form of action, linear in $Q$, takes place only for sets of charges,
satisfying
condition (27). For superpositions with arbitrary charges appear interaction
between different charges.

\bs
\cl{\large {\bf 2. The Cartan tori}}

\bs
In this section we will be interested in degenerated tori, related
with maximal abelian  Cartan tori $T_G$
of the simple compact Lie groups $G$.
We consider their different representations, corresponding homotopical
groups $\pi_1(T_G)$
and topological charges.

The Cartan torus $T_G$ consists of elements
$$
{\bf g} = e^{2\pi i({\bf H}\mbox{\scriptsize{\bm $\phi$}})},\;
{\bf H} = \{H_1,...,H_n\} \in \mf{C}, \; [H_i,H_j] = 0 ,
\en(29)
$$
where $n\mbox{ is a rank of}~G,$
$\mf{C}$ is a maximal commutative Cartan
subalgebra of the Lie algebra $\mf{G}$ of the group $G$ and we assume that
$(H\phi)$ is usual euclidean scalar product.
Due to their commutativity all $H_i$ can be diagonalized simultaneously.
Their eigenvalues are called the weights  ${\bf w}$ and
depend on concrete representation of $G$ and $\mf{C}$ .
The weights $\{{\bf w}_a \}_{\tau}, \,
a = 1,..., p$, belonging to a $p$-dimensional irreducible
representation $\tau(G)$ form a set "of quantum numbers"  of this
representation.

All possible weights  ${\bf w}$ of the simply connected
group $G$ or of the universal covering groups $\tilde G$ of
non-simply connected groups $G$
form a lattice of weights $\mbb{L}^G_{w}$. Its basis can be
choosed by different ways. The basis, most convenient from the view point
of the representation theory, is the so called
basis of fundamental weights
${\bar{\bf w}}_i,\, i = 1,...,n.$
Any weight ${\bf w}$ can be
represented in the form
$$
{\bf w} = \sum\limits^{n}_{1}
n_i \bar{\bf w}_i,
\en(30)
$$
where all $n_i$ are integers. But, in general, not all
$\bar{\bf w}_i$
have a minimal norm. In some cases more convenient is a basis of vectors
with minimal norm. As in a case of all lattices different basises are
related by linear modular
transformations $M$ with $det M = \pm 1$ and integer coefficients \ci{13}.
For transformations, preserving an orientation of the basis, $det M =1.$

The weight lattice ${\mbb L}_{w}$ and related with it root ${\mbb L}_r$
and dual
root ${\mbb L}_v$ lattices are needed for finding $\pi_1(T_G),$ which can
depend on concrete representation $\tau (G).$
The group $\pi_1(T_G)$ can be defined in general form by the
methods of the Lie groups and
algebras \ci{14}, but we will use here more simple and obvious one, analogous
to one used above for tori $T_L$. Since in any irreducible representation
$\tau(G)$ of dimension $p$ one can choose the eigenvectors $|a>$ of
${\bf H}$ as a basis
$$
{\bf H}|a> =  {\bf w}_a |a>,\; a=1,...,p,
\en(31)
$$
then in this basis all $H_i$ and any element ${\bf g}\in T_G$ have diagonal
form
$$
{\bf g}_{\tau}=
diag(e^{2\pi i(w_1 \phi)},...,e^{2\pi i(w_p \phi)})
\en(32)
$$
Here again $(w_a \phi)$ denotes an usual euclidean scalar product of
the vectors
${\bf w}_a$ and {\bm $\phi$}. The main difference of
this form from usual representation of tori $T_L$ type (16)
is that:

1) a dimension of diagonal matrices in (32)
coincides with dimension of $\tau$-representation
$p,$ which is usually larger, than rank of $G$,

2) the set of weights
$\{{\bf w}\}$ has discrete Weyl symmetry, which ensures Weyl invariance
of $T_G$ and results in the next two
properties
$$
\sum_{a=1}^p {\bf w}_a = 0, \quad
g_{ik} = \sum _{a=1}^p w^a_i w^a_k = B_{\tau}\delta_{ik},
\en(33)
$$
where constant $B_{\tau}$ depends on representation. It follows from (33)
that in case of $T_G$ effective metric $g_{ik}$ on the parameter space is
proportional to the euclidean one.
This fact will appear very
important for interaction of the corresponding topological charges and
difference of  properties of non-linear $\sigma$-models on tori $T_L$
and $T_G$ \ci{18}.

From (33) it is obvious, that in this representation all ${\bf g}\in T_G$
are periodic with a lattice of periods ${\mbb L}_{\tau}^{-1}$,
dual (or inverse) to the lattice ${\mbb L}_{\tau}$, which is generated
by weights ${\bf w}_a$ of $\tau$-representation.
The point is that the lattice ${\mbb L}_{\tau}$ can be
some sublattice of lattice ${\mbb L}_{w}$: ${\mbb L}_{w} \supseteq {\mbb
L}_{\tau}.$

A lattice $\mbb{L}_v$, inverse to ${\mbb L}_{w}$, is called a lattice
of dual roots ${\bf r}^v.$
It has its own basis of the dual roots
${\bf r}^v_i \,(i = 1,...,n)$
$$
(r^v_i w_k) = \delta_{ik}.
\en(34)
$$
Thus we see that for simply connected $G$ and representations, containing
weights, belonging to the ${\mbb L}_{w},$ but not to some its sublattice,
$\pi_1(T_G)$ is isomorphic
to the $n$-dimensional
lattice $\mbb{L}_v$ of the dual roots ${\bf r}^v$
(see for example \ci{17}).
All simple compact groups
$G$, except $G = SO(N)$, are simply-connected. For
multiply connected $G$ the group $\pi_1(G)$ is a subgroup of a
finite discrete group -- a center $Z_G$, which belongs to
$\tilde T_G$ -- the maximum torus of a universal covering
group $\tilde G.$ Therefore, for multiply connected $G$ such,
that $G = \tilde G/\pi_1(G)$
$$
\pi_1(T_G) = \pi_1(\tilde T_G) \times \pi_1(G) =
{\mbb L}^G_v\times \pi_1(G)
$$
It means, that $\pi_1(T_G)$ contains ${\mbb L}_v$
as a sublattice. For groups $SO(N)$ $\pi_1(SO(N)) = Z_2$,
and for the adjoint groups $G_{ad} = \tilde G/Z_G$
$$
\pi_1(G_{ad}) = Z_G
$$
For simply-laced  groups $G$ (i.e. for groups of a series
$A_n = SU(n+1), D_n = SO(2n), E = E_{6,7,8}$) ${\mbb L}_v$
coincides with a lattice of all roots ${\mbb L}_r$, which is
a sublattice of a lattice of weights ${\mbb L}_r\subseteq {\mbb L}_{w}.$
The structure of all lattices ${\mbb L}_{w}, {\mbb L}_r, {\mbb L}_{v}$
of simple
compact groups $G$ is known [14]. It is necessary to note, that in
general case they are not any more hypercubic and can belong to four different
series of $A,D,E,Z$ types.
Further we will consider only minimal and adjoint representations.
The weights of minimal representations generate a weight lattice ${\mbb L}_{w},$
thus a lattice of topological charges will be lattice of dual roots ${\mbb
L}_v.$
The weights of adjoint representation are roots and generate root lattice
${\mbb L}_r,$ in this case topological charges will belong to ${\mbb L}_r^{-1}.$

It was shown in \ci{17} that for groups
$G = G_2,E_8,C_n,adA_n,$ $adB_n,adD_n$ and  $adE_{6,7}$ the lattice  of all
possible topological charges ${\mbb L}_\tau\supseteq {\mbb L}_{w},$
and consequently for them there is a possibility of the
topological interpretation of
all "quantum numbers" of these groups. There is also possibility
of a partition
${\mbb L}_\tau^G$ into sets of charges, corresponding to the irreducible
representations of
groups $G'$, for which ${\mbb L}^G_\tau = {\mbb L}^{G'}_{w}.$ For example,
${\mbb L}_{\tau}^{G_2} = {\mbb L}^{A_2}_{w},\; {\mbb L}_{\tau}^{E_8} =
{\mbb L}^{E_8}_{w},\;
{\mbb L}_{\tau}^{C_n} = {\mbb L}_{w}^{C_n}.$ For other groups $G,$ for which
${\mbb L}_{\tau}^{adG} = {\mbb L}^G_{w}$ (for example, $G = A_n,B_n,D_n,E_{6,7}$)
not all
sets of topological charges, answering to some
representation (for example, quark representations of groups
$A_n$ or spinor representations of groups $B_n,D_n$) will correspond
to the exact (single-valued)
representations.

Now let us consider two-dimensional chiral model on tori $T_G.$
The corresponding action ${\cal S}$ is
$$
{\cal S} = \frac{1}{2\alpha} \int d^2x tr_{\tau}({\bf g}^{-1}_{\nu}
{\bf g}_{\nu}) =
\frac{1}{2\alpha} \int d^2x tr_{\tau}( H \phi_{\nu})^2 =
\frac{1}{2\alpha}B_{\tau}\int d^2x ( {\phi}_{\nu})^2,
\eqno (35)
$$
where ${\bf g}=e^{2i\pi ( H \phi)} \in T_G,$
{\bm $\phi$}= $(\phi_1,...,\phi_n),$
$n$ is
a rank of $G,$ $\phi_{\nu} = \partial_{\nu} \phi, \; \nu = 1,2,$
and effective metric $g_{ik}$, generated by weight system
$\{{\bf w}\}_{\tau}$ of $\tau$-representation, from (33)
is used.
It is convenient to introduce in this model a normalized trace
to avoid constant $B_{\tau}$ from effective metric
$$
Tr = \fr{tr_{\tau}}{B_{\tau}}
\en(36)
$$
Then ${\cal S}$ takes a canonical form without coefficient $B_{\tau}.$
The corresponding equations of motion
$$
({\partial}_{\nu})^2 (H \phi) = 0,
\eqno (37)
$$
have classical vortex-type solutions in a region $R>r>a$ analogous to (20)
$$
\mbox{\bm $\phi$}({\bf x})
=\fr{1}{\pi}{\bf Q} \arctan (\fr{y-y_0}{x-x_0}), \;
{\bf Q} \in {\mbb L}_{\tau}^{-1}
\eqno (38)
$$
Just these solutions for groups G, such,
that ${\mbb L}_{\tau}^{-1} \supseteq {\mbb L}_{w}$, can give
the topological interpretation of
all their quantum numbers \cite{17}.
Energy of these vortices is       also
logariphmically divergent
$$
E = 1/(2\alpha)\int (\partial_{\mu} \phi)^2 d^2 x
=\fr{1}{2\alpha} (2\pi) ({\bf Q})^2\ln (R/a),
\en(40)
$$
This gives nonzero logariphmic interaction between vortices with
different non-collinear
topological charges in contrast with a case of tori $T_L$
$$
E = \fr{1}{2\alpha}({\bf Q}_1 {\bf Q}_2){2\pi}
\ln {|{\bf x}_1 - {\bf x}_2|/a}.
\en(41)
$$
This interaction depends on relative orientation of topological charges and
for this reason a geometry of lattices becomes very important.
Energy of $N$-vortex type solutions will have a form similar to (22) with
usual scalar product of topological charges.

Now one can conclude
that properties of $\sigma$-models on the Cartan tori will be different and
depending on lattice ${\mbb L}_{\tau}^{-1}.$
As was mentioned above,
all lattices, connected with compact simple groups, belong to four series
$A,D,E,Z$ and in appropriate scales are integer-valued \ci{20}. How they
determined critical properties of topological phase transitions is
considered in paper \ci{18}(see also \ci{15}).

It is interesting to know which spaces, besides tori, can have homotopical
group $\pi_1$ of lattice type, which must be integer-valued. One must also
note, that there are spaces with  $\pi_1$ having some finite cyclic subgroups,
but we do not discuss them here.

One-dimensional conformal $\sigma$-models on tori $T_G$ can be analysed
as those on tori $T_L.$ The single difference will be that a sum in trace
now goes over all weights of some representation. Since instantons
(or anti-instantons) do not
interact between them, there will be no essential changes in properties
of $\sigma$-models on tori $T_L$ and $T_G$ as long as we will
be interested only unmixed configurations. But they will be different if
one consider properties, connected with all possible configurations.

\bs

\cl{\large{\bf 3. Manifolds with $\pi_2 = {\mbb L}$}}

\bs

Now we pass to consideration of manifolds ${\cal M}$ with homotopical
group $\pi_2$
of lattice type. The more known such manifolds can be divided into two
large classes: 1) homogeneous
spaces of simple compact Lie groups, 2) the Hodge submanifolds of complex
projective spaces ${\mbb C}{\mbb P}^n,$ do not coinciding with ${\cal M}$
from 1). The first ones
are the more interesting from the point of view, accepted in  this paper
because they can have $\pi_2$ group with
topological charges
belonging to the weight lattices (or their sublattices) of the corresponding
groups. For finding out instanton type solutions with topological charges
${\bf q} \in {\mbb L}_r$  the more important are the complex homogeneous
spaces.
The main classification theorem about such manifolds says that all of them
can be considered as bundle spaces over some flag spaces $F$
$$
{\cal M}\stackrel{T}{\lra}F
$$
where a fiber $T$ is a parallelizable space. Moreover $T$ must be some complex
torus if ${\cal M}$ is simply connected or if ${\cal M}$ is kahlerian. If both these
conditions take place, then ${\cal M}$ must be some flag space $F$ \ci{21,22}.
Since
$\pi_2(T)=0,$ the most important from instantons point of view are the flag
spaces $F.$ Among them there are  maximal flag spaces $F_G= G/T_G$ with
$\pi_2(F_G)={\mbb L}_{v}$ \ci{17,23}.

The chiral models on
$F_G$
were considered in papers \cite{23,24},
where it was shown,
that since $\pi_2(G/T_G) = \pi_1({\tilde T}_G) \ne 0$
and on $G/T_G$ there is
complex structure, the appropriate equations have
holomorphic instanton solutions.
But their consideration  of
topological charges was carried out not taking into account vectorial
structure of corresponding group $\pi_2(F_G)$.
The authors have proceeded from the assumption, that appropriate topological
charges are simply integer numbers, corresponding to convolution of some
2-form with different 2-subspaces.
Since $\pi_2(G/T_G) = {\mbb L}_{v}$,
it is possible to consider corresponding topological charges as isovectors.
To see how they can be realized, let us return to $\sigma$-model
on $G/T_G$. Its action has the form \ci{23}
$$
{\cal S}[u] = \frac{1}{2g^2} \int d^2x g_{a\bar b}(u,\bar u)
{\partial}_{\mu} u^a {\partial}_{\mu} \bar {u^b}, \;
a,b = 1,..., p, \; 2p + n = dim G.
\eqno (42)
$$
Here $u^a$ are  complex local coordinates on $G/T_G$,
and the metric $g_{a\bar b}(u,\bar u)$ is a representative
of the 2-forms from class $\varrho$, where {\bm $\varrho$}  is an isovector,
equal to a half-sum of all positive roots of  Lie algebra ${\cal G}$
$$
\mbox{\bm $\varrho$} = \frac{1}{2}\sum_{a> 0} {\bf r}_a =
\sum_{i=1}^n {\bar \omega}_i ,
$$
here ${\bar \omega}_i$ are the fundamental weights of group $G.$
Just this metric is the Einstein one on $G/T_G$ \cite {24,25},
i.e.
$$
g_{a\bar b}(u,\bar u) = k R_{a\bar b}(u,\bar u), \; k = \frac{1}{2},
\eqno (43)
$$
where $R_{a\bar b}(u,\bar u)$ is the Ricci tensor. Such choice of invariant
metric ensures a renormalizability of (42) \cite{24,26}.

On holomorphic fields $u^a(z), \, z\in {\mbb C} = {\mbb R}^2,$ which are
the coordinates
on $F_G,$ action ${\cal S}$ formally coincides (up to coefficient) with
topological invariant
$$
Q = \frac{1}{2} \int d^2x g_{a\bar b}(u,\bar u)
\varepsilon_{\mu\nu}
{\partial}_{\mu} u^a {\partial}_{\nu} \bar {u^b} =
\frac{1}{2} \int g_{a\bar b} du^a \wedge d\bar {u^b}.
\en(44)
$$

As is known \cite{21,24}, all integer-valued 2-forms on $F_G = G/T_G$
$$\Omega = \Omega_{a\bar b} du^a\wedge d\bar {u^b}$$
can be
decomposed (modulo exact forms) on basis of the 2-forms
$\omega_i, \; i = 1,...,n$,
parametrized by fundamental
weights $\bar \omega_i$
$$\Omega = \sum_1^n c_i \omega_i, \; c_i\in {\mbb Z}.
\eqno (45)
$$
Similar expansion with coefficients $c_i = 1$
 has the 2-form
corresponding to the metric $g_{a\bar b}(u,\bar u).$
On the other hand on $G/T_G$ there is a space of 2-cycles
$\gamma$, which is dual to space of the 2-forms.
The 2-cycles are given by linear combinations  of the Shubert  cells
of minimum
dimension: real dimensionality 2 or complex
dimensionality 1,  number of which is equal to a rank of $G$ \ci{21}.
Since $\pi_1(F_G) = 0$ and according to the Gurevich theorem  $\pi_2(F_G) =
H_2(F_G,{\mbb Z})= {\mbb L}_v,$  it is natural to introduce on the space
of 2-cycles
a vectorial structure similar to (9) and to choose a parametrization
of 2-cycles by a lattice ${\mbb L}_{v}$, then its basic cycles $\{\gamma_i\}$
are parametrized
by the simple dual roots ${{\bf r}^v}_i$
$$
\gamma \to \mbox{\bm $\gamma$} = \sum_1^n n_i \gamma_i {\bf r}_i^v,\; n_i\in
{\mbb Z}.
\eqno (46)
$$
A convolution of the 2-forms $\omega_i$ and of  dual to them 2-cycles
$\gamma_k$, i.e.
integral of the 2-form $\omega_i$ over a 2-cycle $\gamma_k$
is equal
$$
(\omega_i \circ \gamma_k) = \int_{\gamma_k} \omega_i = \delta_{ik}
$$
Since  2-form $\varrho$, determining topological charges, is fixed,
the vectorness of space of 2-cycles induces a vectorness of
space of topological charges ${\bf Q}$

$$
{\bf Q} = \sum_{i=1}^n q_i {\bf r}_i^v =
({\bf \varrho} \circ \sum_1^n q_i \gamma_i){\bf r}_i^v,\; q_i\in {\mbb Z}
\eqno (47)
$$

From this it follows that topological charges, answering
to different independent cycles, cannot be simply added and,
in particular, compensate each other.
For evaluations it is useful to note similarity of topological
cell-like structure
of torus
$T_G$ and flag $G/T_G$ in their lowest nontrivial cell
complexes: for a torus $T^n$  it is a bouquet of
circles $T^n = S_1^1\vee ...\vee S_1^1$ \cite{13} and for a flag space it
will be an analog of bouquet of 2-spheres $M = S_1^2\vee ...\vee S_1^2$.

For example, for $G = SU(n+1)$ as local coordinates
on these 2-spheres it is possible to choose the elements of upper, nearest
to a main diagonal,  diagonal line. Then, corresponding
to these 2-spheres instantons are simply instantons of Belavin -
Polyakov \ci{4}. For an action ${\cal S}$  of $n$ instantons with topological
charges $q_1,...,q_n,$ corresponding to $n$ different 2-cycles,
we shall receive
$$
{\cal S} = \frac{1}{2g^2} \sum_1^n |q_i|
\eqno (48)
$$
This expression is similar to that for one-dimensional instantons from (28).

Thus it is shown, that instanton solutions of two-dimensional
chiral
theories on maximal flag spaces $F_G = G/T_G$
can realize isovectorial topological charges,
belonging to $\pi_2(G/T_G) = {\mbb L}_{v}$, and therefore
can realize nearly all quantum numbers of
the appropriate group $G.$ But, as in one-dimensional case, instantons do
not interact with each other. For this reason, only instantons cannot
realize real particles. For their realization one need exitations, interacting
like
vortices, but with Coulomb or Yukava type potential $V(r)$
$$
E = ({\bf q}_1 {\bf q}_2) V(r), \;
V_C(r) = \fr{1}{r}, \; V_Y(r) = \fr{e^{-mr}}{r}
$$
Such type of interaction can be obtained due to gauge fields, which appear
naturally when one consider $T_G$ and $F_G$ as submanifolds in $G$ and
takes into account the fact that in $G$ there is a whole set of such
conjugated submanifolds. Then, if there is no fixing potential, all these
submanifolds will be on equal foot. It means that corresponding
$\sigma$-models fields,
taking their values in these submanifolds, at different points of
physical space can take values in {\it different submanifolds} and
conjugation
transformations will play a role of gauge fields.
A work in this direction is now going on.

\bs

\centerline{\large {\bf 4. 3-dimensional topological exitations}}

\bigskip

Topological exitations of vortex  and instanton types, i.e., connected
with $\pi_1(T_G)$ and $\pi_2(F_G),$ can exist also in three-dimensional space.
Vortex type exitations will form now  closed  or open lines, in a latter
case its energy will be $E \sim L \ln R/a$, where $L$ is a line's length.
Topological exitations
connected with topological charges ${\bf Q} \in \pi_2(G/T_G)$
can exist in the generalized Higgs-Salam-Weinberg models
as 3-dimensional particle-like monopoles. They will have
finite energy if corresponding gauge field would be engaged and
will have divergent energy without gauge field. It is interesting note
that since $\pi_2(F_G) = {\mbb L}_{v}$, which in $G=SU(n)$ case
does not contain weights
of the fundamental quark representations, the latter cannot exist as
topological exitations on $F_G.$ This fact demands additional
factorization of $F_G$ or spontaneous symmetry breaking
if one want to get topological interpretation of quarks
also (see for example \ci{27}).

Here we discuss as an example instantons exitations connected with homotopic
group $\pi_2$ in 3-dimensional
chiral model
on sphere $S^2$ with action
$$
S = \fr{1}{2\alpha} \int d^3x (\partial_{\mu} {\bf n})^2,\,
{\bf n} \in S^2
\en(49)
$$
The corresponding equations have instanton type solution, which we
represent in a form analogous (39)
$$
n_i(x)= \fr{x_i}{r} \theta_r(r-a), \,,
\en(50)
$$
where $\theta_r(r)$ is a regularized step-function.
Its topological charge
$$
Q= \fr{1}{4\pi} \int dS
\en(51)
$$
where integral $\int dS$ is over sphere in internal space, and
energy
$$
E=\fr{1}{2\alpha} \int d^3x (\partial_{\mu} {\bf n})^2 =
\fr{1}{\alpha} 4\pi \int_a^R dr \simeq \fr{4\pi}{\alpha} R.
\en(52)
$$
grows linearly with $R$.
It means that in this model instantons must be confined in neutral pairs
like quarks in QCD and will appear themselves only at small distancies!
All infra-red properties will not depend on their existence. Analogous
properties will have $\sigma$-models on other manifolds having the same
homotopical
 group $\pi_2 = {\mbb Z}$ or of vectorlike type $\pi_2 = {\mbb L}$ as in a case
of $F_G.$

One can show (for example \ci{14}), using exact homotopic sequence for bundle
$G \stackrel{T_G}{\lra}F_G$
and the fact, that $\pi_{i}(T_G) = 0, ~ i > 1$,
$$
\pi_k(G) = \pi_k(G/T_G) = \pi_k(G,T_G), ~k = 3,4...
$$
This  means, that for $k>2$,  the flag space $F_G$ has  the same homotopical
groups as group $G$ itself.
In particular, in a case of 3-dimensional non-linear $\sigma$-models
on flag spaces $F_G$
they, probably,  can have knotted configurations,  connected with
group $\pi_3(F_G)= \pi_3(G) = {\mbb Z},$
similar to those proposed recently for 4-dimensional Yang-Mills theory
in \ci{28}.

All this facts about reach set of topological exitations on Cartan tori
$T_G$ and flag
spaces $F_G$ say
that manifolds $T_G$ and $F_G$ and theories on them
or their gauge generalizations could be considered
as necessary ingredient of any high energy
GUT-type or string-type theories, which try
to obtain the obvious physical
(in particular,  topological) interpretation
of elementary particles.

The author is grateful to Prof. G.Volovik for useful discussion of some
problems, considered in this paper, and to Prof. V.Gurarie for kind
information about literature on complex homogeneous spaces.

The work was supported by RFBR grants 96-02-17331-a and 96-1596861.

\end{document}